\pdfoutput=1

\documentclass{article}
\usepackage[sorting=none, style=numeric-comp]{biblatex}%[sorting=none]
\usepackage{float}
\usepackage{array, caption, tabularx, ragged2e,  booktabs,authblk,pdflscape,tabularray,geometry, setspace,anyfontsize}
\usepackage{graphicx} % Required for inserting images
\usepackage{amssymb}
\usepackage{xcolor}
\usepackage{pdfpages,pgffor}
\usepackage{enumitem}
\usepackage{listings}

\title{The Causal Roadmap and Simulations to Improve the Rigor and Reproducibility of Real-Data Applications}
\author[1]{Nerissa Nance}
\author[1]{Maya L. Petersen}
\author[1]{Mark van der Laan}
\author[1]{Laura B. Balzer}
\affil[1]{University of California, Berkeley}
\date{May 2024}
\begin{document}
\pagecolor{white}
%TC:ignore
\maketitle

\begin{abstract}
The Causal Roadmap outlines a systematic approach to asking and answering questions of cause-and-effect: define the quantity of interest, evaluate needed assumptions, conduct statistical estimation, and carefully interpret results. To protect research integrity, it is essential that the algorithm for statistical estimation and inference be pre-specified prior to conducting any effectiveness analyses. However, it is often unclear which algorithm will perform optimally for the real-data application. 
Instead, there is a temptation to simply implement one's favorite algorithm --- recycling prior code or relying on the default settings of a computing package.
Here, we call for the use of simulations that realistically reflect the application, including key characteristics such as strong confounding and dependent or missing outcomes, to objectively compare candidate estimators and facilitate full specification of the Statistical Analysis Plan.
Such simulations are informed by the Causal Roadmap and conducted after data collection but prior to effect estimation.   
We illustrate with two worked examples. First, in an observational longitudinal study, outcome-blind simulations are used to inform nuisance parameter estimation and variance estimation for longitudinal targeted minimum loss-based estimation (TMLE). Second, in a cluster randomized trial with missing outcomes, treatment-blind simulations are used to examine Type-I error control in Two-Stage TMLE. In both examples, realistic simulations empower us to pre-specify an estimation approach that is expected to have strong finite sample performance and also yield quality-controlled computing code for the actual analysis. Together, this process helps to improve the rigor and reproducibility of our research.
\\
%\emph{Keywords: Causal inference, Causal Roadmap, Pre-specification, Real-world data, Reproducibility, Simulations, TMLE }
\end{abstract}

%TC:endignore
\section{Introduction}
  
Formal frameworks for causal and statistical inference can help researchers to clearly structure and understand the links between their research question, causal model, data, statistical estimation, and results interpretation. Examples of such frameworks include the Causal Roadmap and target trial emulation.\cite{petersen_causal_2014,laan_targeted_2011,hernan_using_2016}
Recent commentaries on epidemiologic training have highlighted the role of such frameworks in asking thoughtful and feasible study questions, particularly amid a proliferation of novel analytic methods which may aid or distract from answering that question.\cite{fox_critical_2020,dang_start_2023} Even after we have specified a well-defined and relevant question, there are many steps to setting up the analysis to answer it. For example, the remaining steps of the Causal Roadmap (hereafter, ``the Roadmap'') are to: (2) specify a causal model reflecting background knowledge and uncertainties; (3) define the causal effect of interest; (4) describe the data available to answer the question; (5) assess identifiability; (6) select a statistical model and estimand; (7) estimate and obtain inference, and (8) interpret results.

Roadmap steps 1-6 set up a statistical estimation problem, reflecting our research question and the real-world challenges of the data. 
Specifically, the Roadmap leads us to a well-defined statistical estimand, which is a function of the observed data distribution, and a realistic statistical model. (Formally, the statistical model is the set of all possible observed data distributions.\cite{laan_targeted_2011})
However, the Roadmap does \emph{not} tell us which algorithm to apply for estimation and inference.
While an algorithm's theoretical properties can narrow the scope of possibilities, it is often unclear \emph{a priori} which approach will perform best in the real-data application. % and, thus, yield the best answer to our research question. 
Instead, there is a tendency to simply apply one's preferred algorithm, using the default settings of a computing package or recycling prior code. Likewise, there is a temptation to try several implementations and pick the implementation yielding the most favorable or logical result. As detailed below, we advocate for the use of realistic simulations to objectively select the algorithm for estimation and inference and pre-specify the Statistical Analysis Plan (SAP).

The SAP delineates key features of the real-data analysis, including the target population, primary outcome, exposure conditions, causal effect of interest, approaches to handling potential inferential threats (e.g., confounding and missing data), statistical estimand, primary/secondary analyses, and sensitivity analyses. While many of these features follow from earlier steps of the Roadmap, the SAP requires us to state the precise implementation of the estimator, including approaches for estimating nuisance parameters and approaches for obtaining inference. (Nuisance parameters are quantities needed to evaluate the statistical estimand, but are not the estimand itself.) Full pre-specification requires more than simply stating the statistical estimand (e.g., the longitudinal G-computation formula) and the general class of estimators (e.g., targeted minimum loss-based estimation [TMLE]). Indeed, the process of pre-specifying an SAP requires us to think critically about different estimation and inferential strategies as well as their expected performance \emph{before} running any analyses to assess causality. As a result, pre-specification of the SAP helps improve transparency and protect against \emph{ad hoc} analyses, which can lead to a ``fishing expedition'' to find the most promising results and inflated Type-I error rates.

Regulatory and funding agencies typically require the SAP to be pre-specified prior to conducting effectiveness analyses in randomized trials.\cite{ich_harmonised_tripartite_guideline_statistical_1998,european_medicines_agency_ich_2020,us_food_and_drug_administration_fda_adjusting_2023,national_institutes_of_health_nih_research_nodate} There is also a growing movement to improve reproducibility and transparency of observational studies through rigorous planning and reporting (e.g., \cite{dang_causal_2023-1,munafo_manifesto_2017,hiemstra_debate-statistical_2019,diaz_sensitivity_2023}).  
Mathur and Fox provide an excellent review on the principles and practices to improve open and reproducible research in epidemiology; in particular, they highlight pre-registration of SAPs for observational studies and code sharing.\cite{mathur_toward_2023}
Notably, Gruber et al. discuss how the Roadmap for Targeted Learning can inform SAP development.\cite{gruber_developing_2023} 
Here, we build on this work by providing guidance and context on \emph{how} to select the approach for statistical estimation and inference for the real-data application.

Our goal is to describe how finite sample simulations, informed by the Roadmap and reflecting the real-data application, can be used to objectively compare estimation strategies and develop a completely pre-specified SAP. For demonstration, we provide two worked examples: (1) an observational study with a time-varying exposure and censoring, and (2) a randomized trial with missing and dependent outcomes. %These case studies are summarized in Tables 1 and 2, and the resulting SAPs are presented in the Appendices. 
We also highlight how this approach naturally leads to fully pre-specified and quality-checked computing code. Thus, our approach has the potential to improve the transparency, reproducibility, and rigor of our analyses aiming to evaluate causal effects. 

Our presentation assumes familiarity with foundational concepts in causal and statistical inference (e.g., causal models, identifiability assumptions, and the G-computation formula). For a review of these concepts and an introduction to the Causal Roadmap, we refer to Petersen and van der Laan\cite{petersen_causal_2014} and Dang et al.\cite{dang_causal_2023}. An overview of the Roadmap for the running examples is provided in Table 1.  Indeed, our worked examples are inspired by real studies and are inherently complex, highlighting the real-world challenges that commonly arise when aiming to infer causality. 
The remainder of the article is organized as follows. First, we outline how simulations are used in epidemiology. Then, we demonstrate the utility of the Roadmap in setting up the statistical estimation problem and designing the simulation study. Next, we describe how to conduct simulations for estimator selection in real-data applications; specifically, we discuss pre-specification of the candidate estimators, data generating process,  performance metrics, and  selection scheme. Finally, we describe the consequences of our approach to improve research transparency and reproducibility. %We close with a brief discussion.

\section{On simulations in epidemiology}

Simulation studies are widely applied in methodological research to evaluate the finite sample properties of existing and recently developed estimators.\cite{morris_using_2019} Since the true value of the target parameter is known, simulations enable us to calculate performance metrics, such as bias %(the average deviation between the point estimate and the true effect) 
and confidence interval coverage% (the proportion of computed confidence intervals that contain the true effect)
. For example, the well-known Kang and Schafer censored data simulations revealed the instability of estimating equation-based methods under data sparsity and inspired suspicion of doubly robust approaches.\cite{kang_demystifying_2007} Subsequent replication of these simulations have highlighted the potential for doubly robust, substitution estimators (e.g., TMLE and collaborative TMLE) to overcome these challenges.\cite{sekhon_propensity-score-based_2011}  More recently, simulations have been used to illustrate the potential advantages and perils of using machine learning in analyses seeking to infer causality.\cite{naimi_challenges_2021,balzer_demystifying_2021,dorie_automated_2019}

Beyond methods evaluation, epidemiology uses simulations in several other settings. Examples include teaching epidemiologic concepts, %evaluating bias in applied analyses, 
evaluating study designs, forecasting disease trajectories, agent-based modeling, addressing transportability, and data pooling
(e.g., \cite{fox_illustrating_2022, anastassopoulou_data-based_2020, althoff_life-expectancy_2019,nianogo_investigating_2019,bykov_comparison_2019,zivich_transportability_nodate,filshtein_proof_2021}). Prior to data collection, simulation studies are commonly implemented to inform the design randomized trials, including power calculations (e.g., \cite{chow_sample_2017,commissioner_modeling_2022,balzer_statistical_2018})  Following data collection and after effect estimation, simulations are also applied in sensitivity analyses, including quantitative bias analysis (e.g., \cite{fox_applying_2021,jayaweera_accounting_2023}). 
However, to the best of our knowledge, there are few published examples of using simulations in the principled comparison and selection of estimators for a real-data analysis \emph{after} data have been collected \emph{but} before any effectiveness analyses are conducted.
Some exceptions include the use of outcome-blind simulations  %(i.e., simulations using the real data but obscuring the true exposure-outcome relationship) 
 to select the primary analysis in a SMART trial,
%Wyss et al.\cite{wyss_synthetic_2022}, who used synthetic data to choose the optimal high-dimensional propensity score analysis in a healthcare database study, 
 to evaluate propensity score estimators within TMLE in a drug safety monitoring study,
and  to compare estimators for longitudinal effects with registry data.\cite{montoya_efficient_nodate, williamson_application_2023,nance_applying_2023}

A pertinent commentary in the \emph{British Medical Journal} called for the broader use of simulations to inform applied data analyses, but also recognized that the implementation and reporting of such studies is the subject of continued debate.\cite{boulesteix_introduction_2020} We aim to help address these and other issues by guiding researchers on the use simulations, informed by the Roadmap and reflecting the real-data application, to aid in the development and full pre-specification of the SAP and corresponding computing code.

\section{Defining the estimation problem with the Causal Roadmap}
\label{sec:design}

As illustrated in Table 1, the first six steps of the Roadmap setup the statistical estimation problem, which is defined by the statistical estimand %(i.e., statistical parameter of interest) 
and the statistical model% (i.e., the set of possible observed data distributions)
.\cite{petersen_causal_2014} Of course, one could specify these elements without the Roadmap. In our experience, however, applying the Roadmap has several strengths relative to other frameworks and the following benefits.\cite{dang_start_2023} Among others, the Roadmap helps clarify the research goals, highlight potential inferential threats, specify the handling of events occurring after the initial exposure or treatment, and facilitate transparent discussions about the plausibility of assumptions.
 Perhaps most crucially, the Roadmap leads to a statistical estimand reflecting our original research question as well as the real-world challenges in the data.
In other words, even if the identifiability assumptions do not hold, the Roadmap guides us to statistical estimand coming as close as possible to the wished-for effect. (The size of the ``causal gap'' can be formally explored in sensitivity analyses and is taken into account during interpretation.\cite{dang_start_2023}) In most cases, our statistical estimand is a complicated function of the observed data distribution and not equal to a single coefficient in a parametric regression. This complexity is needed to generate the most appropriate answer to our research question and often precludes the use of more traditional statistical approaches.
Equally important, the Roadmap highlights that we rarely have the knowledge to support functional form assumptions --- beyond treatment randomization in a trial. Instead, our statistical model is often non-parametric or semi-parametric, and we need to harness machine learning during estimation to avoid unsubstantiated assumptions. 

As a concrete example, consider a study aiming to evaluate the effect of a time-varying and non-randomized exposure: sustained use of sodium-glucose cotransporter 2 (SGLT2) inhibitors on the onset of renal disease among patients with diabetes. As shown in Table 1, application of the Roadmap highlights the potential for bias and misleading inference due to confounding, censoring, and  practical violations of the positivity assumption, occurring when there is insufficient variability in the exposure within confounder strata.\cite{petersen_diagnosing_2012,rudolph_when_2022}  
These inferential threats can be particularly fraught in settings with longitudinal exposures; the longer follow-up time, the more potential there is for time-dependent confounding, right-censoring, and lower support for the longitudinal exposures of interest. Given these challenges, the Roadmap leads to a complex statistical estimand: a contrast of the iterated conditional expectation expression of the longitudinal G-computation formula (Appendix-A).\cite{robins_new_1986,bang_doubly_2005}  Importantly, the Roadmap also leads to a non-parametric statistical model without functional form assumptions. Altogether, the Roadmap narrows the scope of possible estimators to algorithms that can handle time-dependent confounding, right-censoring, and poor data support as well as harness machine learning to avoid unsubstantiated modeling assumptions. For this setting, common approaches include singly robust estimators, such as inverse probability weighting and G-computation, as well as doubly robust alternatives, such as augmented inverse probability weighting and TMLE.\cite{horvitz_generalization_1952,rosenbaum_central_1983,robins_new_1986,robins_estimation_1994,laan_targeted_2011} Each has statistical properties that may lend themselves (or not) to a specific analysis. As described below, we can use simulations, informed by the Roadmap, to chose the estimator expected to perform best in the actual analysis.

As a second example, consider the SEARCH-Youth study, a cluster randomized trial to evaluate the effect of a multi-component intervention on viral suppression among youth with HIV in East Africa.\cite{ruel_multilevel_2023} As shown in Table 1, the Roadmap highlights the impacts of randomizing the treatment to health clinics (instead of individuals) and missing data. Specifically, each Roadmap step reflects the dependence between participants within clinics as well as the potential biases from the missing data--equivalent to time-dependent confounding.
Again, %given these real-world considerations, 
the Roadmap leads to a complex statistical estimand: a contrast of clinical-level summary measures, each accounting for baseline and post-baseline causes of measurement and outcomes (Appendix-B).\cite{balzer_two-stage_2021, benitez_defining_2023,nugent_blurring_2023} The Roadmap also leads us to a semi-parametric statistical model, only reflecting our knowledge of treatment randomization. Here, the Roadmap narrows the set of possible estimators to those that flexibly handle dependent and missing data --- specifically, approaches allowing the missingness mechanism to vary by cluster.\cite{balzer_two-stage_2021} In Two-Stage TMLE, for example, we first estimate a summary measure accounting for missing data in cluster separately and then evaluate the intervention effect on those cluster-level summaries. 
Additionally, as common in cluster randomized trials,\cite{kahan_increased_2016} few clinics were randomized, specifically 28,  in SEARCH-Youth. Therefore, we also need an estimation and inferential approach that performs well with few independent units. Again, simulations can aid in the formal evaluation of alternatives and pre-specification of the primary analysis.

\section{Simulations to inform the real-data analysis}

We now detail how simulations, informed by the Roadmap and reflecting the real-data application, can aid in objectively selecting and appropriately implementing the estimation and inferential approach expected to perform best in the real-data analysis. To do so, we need to pre-specify the candidate estimators, data generating process for the simulations,  performance metrics and  selection process.

\subsection{Choosing the candidate estimators}%
\label{sec:spec}

It is essential to choose candidates targeting the statistical estimand of interest. This may seem obvious, but without careful consideration, we could end-up comparing estimators of marginal versus conditional effects, especially in hierarchical data settings.\cite{hubbard_gee_2010,fitzmaurice_estimation_2008,petersen_targeted_2014,benitez_defining_2023} As previously discussed, following the Roadmap narrows the set of candidate algorithms to those targeting the statistical estimand. This set can further be narrowed by considering the asymptotic properties of the estimators (e.g., efficiency, double robustness). Even if we settle on a single class of estimators, such as TMLE, there are still many decisions before the SAP is fully specified.

We must decide \emph{how} to estimate nuisance parameters. 
In doubly robust estimators, for example, nuisance parameters typically include the outcome regressions (i.e., the conditional expectation of the outcome given  past exposure/measurement and covariates) and propensity scores (i.e., the conditional probability of exposure/measurement given the past). 
To respect our statistical model, 
machine learning is often required for flexible, data-adaptive estimation of nuisance parameters.
However, application of  machine learning requires additional choices. For example, in the ensemble algorithm Super Learner, we need to specify the candidate learners (including their tuning parameters), the cross-validation scheme, and the loss function.\cite{laan_super_2007,phillips_practical_2023} 
After obtaining initial estimates of the nuisance parameters, there may be additional decisions. For example, with practical positivity violations, we can decide to truncate the estimated propensity scores at various levels.\cite{petersen_diagnosing_2012,gruber_data-adaptive_2022}  
Finally, there are a variety of options for statistical inference. 
For TMLE, for example, some approaches for variance estimation are the non-parametric bootstrap, standard or cross-validated estimates of influence curve, 
plug-in estimation of the variance, or other doubly robust options.\cite{laan_targeted_2011,balzer_adaptive_2016, benkeser_doubly_2017,tran_robust_2023,laan_targeted_2018} 

For our running examples, Table 2 provides an overview of the candidate approaches that were pre-specified for objective comparison in simulations. In the observational study, the candidate algorithms were longitudinal TMLE with various implementations. For nuisance parameter estimation, Super Learner with and without covariate screening as well as bounded or unbounded estimates of the propensity score were considered. For statistical inference, candidates included Wald-Type 95\% confidence intervals with variance estimated by the influence curve or the non-parametric bootstrap. For the cluster randomized trial, the candidate algorithms were limited to Two-Stage TMLE with the following specifications. For estimation of the cluster-level endpoints accounting for missing outcomes, candidates were  TMLE using Super Learner, TMLE using parametric regressions, and the empirical mean among those measured. For estimation of the intervention effect, candidates were TMLE with various approaches to covariate adjustment for precision gains.\cite{balzer_adaptive_2016,balzer_adaptive_2023}. Finally, candidates for variance estimation included standard or cross-validated estimates of the influence curve. We now discuss how to define the data generating process for the simulation to formally evaluate the performance of these candidates.% in scenarios reflecting the application. 

\subsection{Defining the data generation process for the simulation} 
\label{DGP}

Thus far, the Roadmap has aided in defining the statistical estimation problem and specifying the set of candidate algorithms for estimation and inference. Simulations reflecting the real-data application can facilitate objective comparison and selection between these candidates \emph{if} we choose a data generating process that is close to the real one. Concretely, application of the Roadmap highlighted several potential biases and inferential threats in the running examples (Table 1). For the observational study, we need to design a simulation with, at minimum, the same exposure/confounder/censoring structure and, therefore, the same practical positivity challenges as the real-data. For the randomized trial, we need to design a simulation with, at minimum, the same number of clusters, a similar distribution of participants per cluster, and plausible missing data mechanism as the real-data. Given these specifications, several options exist.% for generating the data. 

Monte Carlo simulations, where we repeatedly sample from a known data generating process, are common and traditionally employ parametric models for data generation.\cite{rubinstein_simulation_2017}  
Such parametric models often fail to reflect the complexities of the real-data, especially in longitudinal or clustered data settings.  
Considering the limitations of fully parametric simulations, plasmode simulations have gained popularity and may be particularly useful for our focus: estimator selection \emph{after} data have been collected \emph{but before} effect estimation. Plasmode simulations, as defined here, encompass a range of semi-parametric methods that sample partially from the empirical data distribution, while allowing for some user-specification.\cite{petersen_diagnosing_2012, schreck_statistical_2023} 

There are various types of plasmode simulations. 
In ``outcome-blind'' plasmode simulations, we preserve the relationships between the baseline covariates, while simulating the outcome (and other variables) through parametric or semi-parametric methods (e.g., \cite{montoya_efficient_nodate,benkeser_improving_2021,williamson_application_2023,wyss_synthetic_2022}). In these simulations, the value of the (simulated) effect is known, but we remain blinded to the true exposure-outcome relationship. As described in Table 2 and Appendix-A, outcome-blind simulations 
were conducted in the observational study by resampling the baseline covariates  from the empirical distribution and then applying highly adaptive LASSO to simulate the longitudinal exposures, censoring, time-varying covariates, and outcome.\cite{benkeser_highly_2016} This approach preserves the complex relationships between baseline covariates, while generating the remaining variables to reflect challenges in the real-data application (e.g., poor data support due to the rare exposure, long-term follow-up, and strong confounding).

``Treatment-blind'' simulations are another plasmode simulation technique where  the covariate-outcome data are preserved but the treatment indicator is randomly permuted.\cite{balzer_adaptive_2023}  As detailed below, treatment-blind simulations are particularly relevant for evaluating Type-I error control, because the null hypothesis is true by design. As outlined in Table 2 and Appendix-B,  such simulations were implemented in the trial example by randomly shuffling the treatment indicator and imposing missingness on outcomes through a measurement indicator, which was generated by an independent statistician and as a function of the baseline cluster-level and individual-level covariates, the permuted treatment indicator, and time-varying covariates. This simulation approach preserves the covariates and underlying outcome, while facilitating a rigorous comparison of alternative approaches 
and their potential to reduce bias due to differential outcome measurement and improve efficiency through covariate adjustment, as described next.

\subsection{Specifying the performance metrics and selection approach}%
\label{sec:performace}

Once we have the set of candidate estimators and data generating process for the simulations, we need to pre-specify the performance metrics and process to objectively compare the candidates. In Table 3, we review some common  metrics, such as the bias and variance of the point estimates as well as 95\% confidence interval coverage (i.e., the proportion of calculated confidence intervals that contain the true effect).
To compare estimators in a way that is agnostic to the variance estimator and evaluate the extent to which an estimator's bias is negligible, we can use ``Oracle coverage'', where the 95\% confidence intervals are calculated using the variance of the point estimates across the simulation iterations, instead of the estimated variance.  
In simulations to inform randomized trials, common metrics include statistical power (i.e., the proportion of times the false null hypothesis is rejected) and Type-I error control (i.e., the proportion of times the true null hypothesis is rejected). We may additionally be interested in estimating the potential savings in sample size to achieve the same power.\cite{vaart_asymptotic_2000,benkeser_improving_2021,balzer_adaptive_2023} Finally, we pre-specify the selection process for objectively choosing the best-performing candidate and, thereby, the primary analytic approach.

For the running examples, Table 2 provides the performance metrics, selection process, and final estimator. In both studies, selection was a two-step process, implemented in \texttt{R}, and with 1000 simulation iterations. In the observational study, the optimal approach for nuisance parameter estimation was first selected to minimize the empirical variance but preserve Oracle coverage. Then given this choice, the optimal approach for inference was selected to minimize the variance estimate but preserve 95\% confidence interval coverage. In the randomized trial, the optimal approach for estimating the cluster-level endpoints accounting for missing outcomes was based on attaining nominal confidence interval coverage for those endpoints and Type-I error control for the intervention effect. Then given this choice, the optimal approach for estimation and inference for the intervention effect was selected to maximize precision without sacrificing Type-I error. Estimation approaches with good, but not optimal, performance were then pre-specified as sensitivity analyses.

\section{Fostering transparent and reproducible research}

Informed by the Roadmap, we have conducted a simulation study, reflecting the real-data application and facilitating objective comparison of various approaches for estimation and inference. Specifically, we pre-specified our candidates, the data generating process, the performance measures, and the selection scheme. With this simulation study, we have a responsible and ``hands-off'' approach to selecting the best estimator and, thus, the primary analytic approach for our real-data application. The corresponding SAPs for the running examples are given in the Appendices and include the design and results of the simulation study. As illustrated in Table 1, interpretation of the study results must account for the statistical assumptions of the selected estimator as well as the plausibility of the identifiability assumptions. %Notably, the pre-selection of estimation strategies is subject to the same restrictions or limitations; these must be transparently reported as well.%The SAPS also include sensitivity analyses that are defined by alternative approaches, which performed well but not optimally (e.g., preserved 95\% confidence interval coverage but had lower precision). 
Altogether, our approach facilitates objective selection and implementation of the best analysis to answer our research question, while protecting research integrity by ensuring we remain blinded to the true causal effect. Importantly, our approach is in-line with regulatory guidelines to update the SAP based on a blinded review of the data.\cite{ich_harmonised_tripartite_guideline_statistical_1998} 

Our proposed process has several consequences for improving research transparency and reproducibility. First, the simulation leads to a fully pre-specified SAP, where analytic decisions are clearly stated and can be critically evaluated. Second, conducting the simulation requires implementing all candidate estimators in computing code. Thus, these simulations serve as an invaluable tool for debugging code and identifying potential issues (e.g., lack of convergence due to rare outcomes) that may arise in the real-data application. Uploading the computing code and results from both the simulation study and real-data analysis to an online repository, such as GitHub, and including detailed explanation through a markup language further improves reproducibility, trust, and open science.

\section{Discussion}

For real-data analyses, we have outlined how simulations can guide the objective selection of the optimal approach for estimation and inference and, thus, full pre-specification of the SAP. Anchored on the Roadmap, these simulations are designed with our research question at the forefront and to explore the primary concerns of the real-data application. 
The results of the simulation may ultimately reveal that it is not feasible to reliably estimate the statistical estimand of interest. In such cases, we may need to return to early steps of the Roadmap and modify the research question and causal estimand to accommodate the limitations in the real-data.\cite{petersen_diagnosing_2012} %Additionally, a valuable future direction for this work will be to incorporate approaches for partial identification into the Roadmap, the corresponding simulations, and the resulting SAP.\cite{tamer_partial_2010,robins_identification_1996,cole_nonparametric_2019,breskin_using_2019,ghassami_partial_2023} 

Further guidance %on simulations to inform applied analyses 
is needed, and our presentation has limitations. First, we focused on plasmode simulations and did not cover alternative approaches, which might be needed if the real-data are not available or only partially available. Additionally, we emphasized the need to emulate the real-data  closely, but did not discuss how to assess the quality of the emulation. It is worth noting that creating simulated data that are ``too close'' to the true distribution can inspire fears of %bias due to 
data ``snooping'';\cite{fisher_treatment_2020}  pre-specification and code-sharing can help alleviate these fears.
Third, our presentation did not cover practical implementation, such as how to vary simulation parameters, determine the number of iterations, and parallelize; we refer to Morris et al.\cite{morris_using_2019} 
for an excellent overview of these and other considerations.  Additional practical details on the projects inspiring our running examples are available in Nance et al.\cite{nance_applying_2023} and Balzer et al.\cite{balzer_statistical_2022}.
Fourth, while our worked examples incorporated common challenges, including confounding, dependence, and missing data, they did not cover other concerns like generalizability,  transportability, and partial identification in detail.\cite{robins_analysis_1989,manski_nonparametric_1990,stuart_use_2011,bareinboim_general_2013,swanson_partial_2018} %However, these can also be incorporated throughout.\cite{petersen_causal_2014,balzer_all_2017,rudolph_transporting_2021} 
Finally, we have presented a two-step process for estimator selection and implementation: (1) conduct a realistic simulation study to objectively compare pre-specified estimators according pre-specified metrics, and (2) implement the optimal estimator (as defined by the simulation study) for the real-data analysis. Alternative approaches, such as auto-TMLE, are being developed  to dynamically evaluate estimators in simulations and implement the optimal estimator in a single step.\cite{shortreed_automated_2020,benkeser_nonparametric_2020}

Altogether, we believe simulations, anchored on the Roadmap, are an indispensable and under-utilized tool for the objective comparison of approaches for estimation and inference in real-data applications. They are a crucial alternative to the status quo: naively applying a preferred algorithm or trying several algorithms and selecting the ``best'' in an \emph{ad hoc} manner. Instead, our  approach provides a formal framework for comparative assessment of alternative strategies for estimation and inference, pre-specification of the corresponding SAP, and generating quality-controlled computing code. Our approach strives to improve the transparency, rigor, and reproducibility of real-data analyses in epidemiology and beyond.

%TC:ignore
\section{Acknowledgements}
This work was supported, in part, by a philanthropic gift from the Novo Nordisk corporation to the University of California, Berkeley for the Joint Initiative for Causal Inference (JICI). The funders had no role in the conceptualization or writing of the manuscript. We gratefully acknowledge the JICI collaborators who inspired the diabetes data example as well as the SEARCH-Youth study team and participants who inspired the randomized trial example. We also thank Drs. Edward Bein and Andrew Mertens for their generous feedback on the manuscript draft.

\clearpage

\captionof{table}{Overview of the Causal Roadmap for the two running examples}
\begin{flushleft}
\begin{table}
\small
\noindent\setlength\tabcolsep{4pt}%
%\begin{tabularx}{\linewidth}{l|c*{3}{>{\RaggedRight\arraybackslash}X}}
% \fontsize{9pt}{9pt}{9pt}\selectfont
%\newcolumntype{b}{>{\hsize=0.5\hsize}X}
\begin{tabularx}{16cm}{X|*{3}{>{\RaggedRight \arraybackslash}X}} 
\hline
\bf{Causal Roadmap steps} 
%\begin{itemize} \item SAP considerations
%\end{itemize} 
& \bf{Observational study example} & \bf{Randomized trial example}  \\
\hline
(1) Research question \begin{itemize}[itemsep=0pt,parsep=0pt,topsep=0pt, leftmargin=*]
    \item Specify the primary exposure(s), outcome, and target population \end{itemize}&
What is the effect of SGLT2 inhibitor use on the risk of renal disease onset among patients with diabetes in an integrated healthcare system in the United States?& 
What is the effect of the multi-component, SEARCH-Youth intervention on viral suppression among 15-24 year-olds with HIV in rural Kenya and Uganda?\cite{ruel_multilevel_2023}
\\
\hline
(2) Causal model  \begin{itemize}[itemsep=0pt,parsep=0pt,topsep=0pt, leftmargin=*]
    \item  Describe the confounding structure, missing data mechanism, (in)dependence structure, \ldots \end{itemize}&  
Longitudinal model including time-varying confounders and factors influencing censoring& 
Multi-level model reflecting that the intervention is randomized at the clinic-level, but outcomes are at the individual-level and subject to missingness 
\\
\hline 
(3) Causal effect \begin{itemize}[itemsep=0pt,parsep=0pt,topsep=0pt, leftmargin=*]
    \item  Using counterfactuals, specify the target effect and scale of interest %(data-adaptive parameter or not) 
    \end{itemize}&
Causal risk difference: $\mathbb{E}[Y_{\bar{a}=1,\bar{c}=0}(t)] - \mathbb{E}[Y_{\bar{a}=0,\bar{c}=0}(t)]$ with  $Y_{\bar{a},\bar{c}}(t)$ as the counterfactual outcome at time $t$ under SLGT2 use $\bar{A}=\bar{a}$ and no censoring $\bar{C}=0$ throughout follow-up
&
 Sample prevalence ratio: $\frac{1/N \sum_i Y_i^c(1)}{1/N \sum_i Y_i^c(0)}$ with $Y_i^c(a)$ as the counterfactual \% with viral suppression in clinic $i=\{1,\ldots,N\}$ under treatment-level $A^c=a^c$ and complete outcome measurement
 \\
\hline
(4) Observed data \begin{itemize}[itemsep=0pt,parsep=0pt,topsep=0pt, leftmargin=*] 
    \item  Describe the data that have been or will be observed, include the exposure(s), outcome(s), $\ldots$ \end{itemize} & 
Longitudinal data structure: history of covariates $\bar{L}(t)$, exposure $\bar{A}(t)$, censoring $\bar{C}(t)$, and outcomes $\bar{Y}(t)$ through time $t$ & 
Multi-level data structure:  cluster- and individual-level baseline covariates $(E^c,W)$,  cluster-level exposure $A^c$, and individual-level, post-baseline covariates $M$,  measurement indicator $\Delta$, and outcome $Y$
\\
\hline
(5) Identifiability  \begin{itemize}[itemsep=0pt,parsep=0pt,topsep=0pt, leftmargin=*]
    \item  Evaluate the assumptions needed to express the causal effect as statistical estimand
    %Positivity; unmeasured confounding 
    \end{itemize}& No unmeasured confounding, sufficient data support (i.e. no practical positivity violations), and censoring-at-random
    % Lack of data support %for causal effect may   causing practical positivity violations 
    & 
    Within each clinic and values of adjustment variables, individual-level outcomes are missing-at-random and there is a positive probability of measurement
\\
\hline
(6) Define the statistical estimation problem  \begin{itemize}[itemsep=0pt,parsep=0pt,topsep=0pt, leftmargin=*]
    \item Specify the statistical estimand and model% Must have an estimator targeting this estimand and respecting the statistical model  
    \end{itemize}& 
    %LBB: For next version - why is your stat model semip vs. nonp? Also can you add your equation
    The statistical estimand is the iterated conditional expectation expression of the longitudinal g-formula.\cite{robins_new_1986,bang_doubly_2005} The statistical model is non-parametric.$^*$ & 

 Within each clinic, the cluster-level endpoint, accounting for missingness, is $Y^c=\mathbb{E}[\mathbb{E}(Y|\Delta=1,W,M)]$. The statistical estimand to evaluate the intervention effect is $\Psi=\frac{1/N \sum_i \mathbb{E}(Y^c|A^c=1, E^c_i)}{ 1/N \sum_i\mathbb{E}(Y^c|A^c=0, E^c_i)}$.\cite{balzer_two-stage_2021} The statistical model is semi-parametric.$^*$
\\
\hline
(7) Estimate and obtain inference \begin{itemize}[itemsep=0pt,parsep=0pt,topsep=0pt, leftmargin=*]
    \item In a pre-specified way, chose and implement an estimator  \end{itemize}&
    Outcome-blind simulations to objectively select between alternative implementations of longitudinal TMLE; see Table 2.\cite{laan_targeted_2018,petersen_targeted_2014} &
    Treatment-blind simulations to objectively select between alternative implementations of Two-Stage TMLE; see Table 2.\cite{balzer_two-stage_2021} \\
   % L-TMLE with discrete superlearner for nuisance parameter estimation and empirical influence curve-based variance estimation\cite{laan_targeted_2018,petersen_targeted_2014,lendle_ltmle_2017} &
   % Two-Stage TMLE to (1) flexibly estimate the cluster-level endpoint $Y^c$ in each clinic, (2) efficiently evaluate the cluster-level intervention effect $\Psi$, and (3) obtain inference via the estimated, cluster-level influence curve.$^*$\cite{balzer_two-stage_2021}   \\
\hline
(8) Interpret \begin{itemize}[itemsep=0pt,parsep=0pt,topsep=0pt, leftmargin=*]
    \item State the results in light of the causal and statistical assumptions. \end{itemize} & 
    After flexibly accounting for measured time-varying confounders, two years of continuous use of SGLT2 inhibitors was associated with a 5\% (95\%CI: 2.75-7.25\%) decrease in the risk of renal disease onset. & 
    After flexibly accounting for missing outcomes and clustering,  SEARCH-Youth  increased viral suppression among adolescents and young adults with HIV by 10\% (risk ratio=1.10, 95\%CI: 1.03-1.16).\cite{ruel_multilevel_2023}\\
\hline 
\end{tabularx}
%\end{tabular}
$^*$See the Statistical Analysis Plan (SAP) in the Supplementary Materials for additional details. 
\end{table}
\end{flushleft}

\clearpage

\captionof{table}{Overview of simulation setup and results for the two running examples.}
\begin{center}
\begin{table}
\small
\noindent\setlength\tabcolsep{4pt}%
%\begin{tabularx}{\linewidth}{l|c*{3}{>{\RaggedRight\arraybackslash}X}}
\begin{tabularx}{16cm}{X*{3}{>{\RaggedRight \arraybackslash}X}} 
\hline
\bf{Simulation specifications \& results}& \bf{Observational study example} & \bf{Randomized trial example}  \\
\hline
Statistical estimation problem$^a$  &
The longitudinal G-computation formula under practical positivity violations with a long-term rare exposure, right-censoring, and rare outcome + non-parametric statistical model &
Two-Stage estimand with (1) differential missingness of individual-level outcomes; (2) few clusters randomized, and (3) the cluster as the independent unit + semi-parametric statistical model \\
\hline
Estimators compared$^b$ & 
Longitudinal TMLE with alternative approaches for\begin{itemize}[itemsep=0pt,parsep=0pt,topsep=0pt, leftmargin=*]
\item Nuisance parameter estimation using Super Learner with/without covariate pre-screening and with/without truncation of the estimated propensity scores
\item Variance estimation: influence curve vs. non-parametric bootstrap
\end{itemize}
& 
Two-Stage TMLE with alternative approaches to 
\begin{enumerate}[itemsep=0pt,parsep=0pt,topsep=0pt, leftmargin=*]
  \item[(1)] account for missing individual-level outcomes: the empirical mean, TMLE with main terms regression, or TMLE with Super Learner
  \item[(2)] adaptively adjust to improve precision: Adaptive Pre-specification with a limited vs. expanded adjustment set 
  \item[(3)] obtain inference with standard or cross-validated estimates of the influence curve \cite{balzer_adaptive_2016, balzer_two-stage_2021, balzer_adaptive_2023}
\end{enumerate} 
\\
\hline
Data generation process&
Outcome-blind simulations preserving the baseline covariate structure, while simulating exposure-censoring variables with positivity challenges and a synthetic outcome with a similar marginal distribution as the real-data&
Treatment-blind simulations preserving the covariate-outcome data structure but randomly permuting the treatment indicator + generation of outcome measurement indicators by an independent statistician 
\\
\hline
Performance metrics and selection approach$^b$ &
Over 1000 iterations,
\begin{enumerate}[itemsep=0pt,parsep=0pt,topsep=0pt, leftmargin=*]
\item[(i)] Select the approach for nuisance parameter estimation that minimizes empirical variance and preserves Oracle coverage
\item[(ii)] Given (i), select the approach for variance estimation that minimizes the estimated variance and preserves 95\% confidence interval coverage.
\end{enumerate}
%Bias, variance, coverage over 1,000 iterations
& 
Over 1000 iterations,
\begin{enumerate}[itemsep=0pt,parsep=0pt,topsep=0pt, leftmargin=*]
\item[(i)] Select the approach for the cluster-level endpoints resulting in nominal confidence interval coverage for those endpoints and Type-I error control for the overall effect. 
\item[(ii)] Given (i), select the  approach for effect estimation and variance estimation resulting in optimal Type-I error control
\end{enumerate}
\\
\hline

Resulting primary analysis$^b$ & 
 % From Roadmap table - repetitive moving here.
 % L-TMLE with discrete superlearner for nuisance parameter estimation and empirical influence curve-based variance estimation\cite{laan_targeted_2018,petersen_targeted_2014,lendle_ltmle_2017} &
%    Two-Stage TMLE to (1) flexibly estimate the cluster-level endpoint $Y^c$ in each clinic, (2) efficiently evaluate the cluster-level intervention effect $\Psi$, and (3) obtain inference via the estimated, cluster-level influence curve.$^*$\cite{balzer_two-stage_2021} 

Longitudinal TMLE with Super Learner %(with candidate algorithms GLM, penalized regression, generalized additive models, gradient boosting models, and neural networks) 
with algorithm pre-screening and without propensity score truncation, with influence curve-based variance estimation.\cite{laan_targeted_2018,petersen_targeted_2014,lendle_ltmle_2017}
& (1) TMLE with Super Learner  to flexibly estimate the cluster-level endpoints, (2) TMLE with Adaptive Pre-specification with limited candidates to efficiently estimate the intervention effect; (3) variance estimation with the cluster-level, influence curve
\\
\hline
\end{tabularx}
$^a$See Table 1 for details. $^b$See the SAPs in the Supplementary Materials for details, including the precise descriptions of the candidate estimators (e.g., the library, loss-function, and cross-validation scheme for Super Learner), performance metrics, and the primary analysis.
%$^2$For Super Learner, we  pre-specified the candidates algorithms as generalized linear models (GLMs), generalized additive models, and the mean; 10-fold cross-validation, and the non-negative least squares loss.\\
%$^3$ We pre-specified two versions of Adaptive Pre-specification: ``limited'' with candidate algorithms as working GLMs adjusting for at most 1 covariate and ``expanded'' with candidates as working GLMs adjusting for a single covariate, working GLMs adjusting for multiple covariates, penalized regression, and the unadjusted estimator. Both implementations used leave-one-out cross-validation, and the estimated influence-curve-squared as loss function.\cite{balzer_adaptive_2016, balzer_adaptive_2023}
\end{table} 
\end{center}

\clearpage

\captionof{table}{Examples of performance metrics to use in simulations to inform real-data analyses.}
\begin{center}
\begin{table}
\small
\noindent\setlength\tabcolsep{4pt}%
%\begin{tabularx}{\linewidth}{l|c*{3}{>{\RaggedRight\arraybackslash}X}}
 \begin{tabularx}{16cm}{X*{3}{>{\RaggedRight \arraybackslash}X}} 
\hline
 \bf{Performance metric}& \bf{Calculation} & \bf{Relevance}\\
\hline 
Bias (point estimates) & %The difference between the expected value of the estimate and the truth 
 The average difference between the point estimate and the target effect across the simulation iterations
 % & Ensuring our point estimates is not far from what you are trying to estimate\\%&
 % $E[\hat{\theta}]-\theta$
 & What is the accuracy of the estimator? \\
\hline 
Variance (point estimates) & The variance of the point estimates across the simulation iterations & How precise is the estimator? \\
\hline 
Mean-squared error & The average of the squared differences between the point estimate and target effect (equivalent to Bias$^2 +$variance) &  What is the variability of the estimator around the target effect?\cite{fox_applying_2021} (Akin to asking how is the estimator balancing bias and variance?) \\
  %Var(\hat{\theta}) =  \frac{\sum_1^k %\hat{\theta}_k - \bar{\theta}}{k-1} 
\hline
Bias-variance ratio & Ratio between the bias and variance  & Is the estimator's bias disappearing at a fast enough rate relative to its variance?\\
 \hline 
Variance to estimated variance ratio & Ratio between the variance of the point estimates and the average variance estimate & Is the variance being over- or under-estimated? \\
   \hline 
  Oracle coverage & The proportion of 95\% confidence intervals, calculated with the variance of the point estimates, containing the target effect & Is the bias in the point estimates impacting the estimator's  confidence interval coverage? \\
  \hline
Confidence interval coverage& The proportion of 95\% confidence intervals, each calculated with the estimated variance, containing the target effect & Does the approach for estimation and inference result in valid inference? \\
 \hline 
Power& The proportion of simulation iterations where the true null hypothesis is rejected & Will the approach for estimation and inference identify an effect when it exists?\\
 \hline 
   Type I error& The proportion of simulation iterations where the true null hypothesis is rejected  & Will the approach for estimation and inference lead to the incorrect conclusion of an effect when none exists? \\
     \hline 
% Mean squared error   & & \\%& $E[(\hat{\theta}-\theta)^2]$ 
%  \hline 
 %Sample size savings    & 1- ratio of MSE &  measure of sample sized saved when using a more efficient estimator for the same power\\

\end{tabularx}
\end{table} 
\end{center}

\clearpage
\printbibliography
\clearpage



\section*{Supplementary material}

# Supplementary Material A

# Causal effects of SGLT2 inhibitor use on renal disease in a population of adults with diabetes

EXAMPLE STATISTICAL ANALYSIS PLAN
v1.1

This analysis plan is inspired by work from the Joint Initiative for Causal Inference (JICI) and follows the Roadmap for Causal and Statistical Inference.[1] The analysis that inspired this work can be found on arxiv (2310.03235).[2] Part of the example analytic approach also inspired by work with Dr. Romain Neugebauer at Kaiser Permanente Northern California Division of Research.



## Research question

What is the effect of sustained exposure to SGLT2 inhibitors (a class of diabetes drug) over time on chronic kidney disease (CKD) among a cohort of patients with diabetes in an integrated health system in the US?

## Causal model

We define the following endogenous nodes for time t=1,...,K+1:
X = {L(t), A(t)}
where:
K+1 is the maximum follow up time of two years. (We explicitly define the time discretization in the Observed Data section below.)

A(t)=($A_1$(t), C(t))

- $A_1$(t) is SGLT2i use at time t
- C(t) is censoring indicator by time t

L(t)=($L_1$(t),Y(t))
- $L_1$(0) defines the baseline covariates are at t=0 and includes age, sex, education, income, neighborhood deprivation index
- $L_1$(t) defines the time-varying covariates at time t and includes insulin use, cardiovascular disease (e.g., ischemic heart disease, myocardial infarction), hypertension
- Y(t) is the outcome, chronic kidney disease, by time t

And exogenous error:
$U = (U_{A(t)}, U_{L(t)})$ for all t

Structural equations are as follows:

$$L(t) = f_{L(t)}(\bar{A}(t-1), \bar{L}(t-1), U_{L(t)}), t = 1,..., K+1$$

$$A(t) = f_{A(t)}(\bar{A}(t-1), \bar{L}(t), U_{A(t)}), \ t = 1,..., K$$

where the overbar denotes the history of the variable

## Causal effect of interest

*The counterfactual difference in the risk of CKD between exposure to SGLT2 vs no exposure to SGLT2 over time*

(a) <u>Treatment regimes of interest</u>

Our treatment regimes of interest correspond to sustained SGLT2 use ($\bar{a}_1 = 1$) throughout follow-up, compared to sustained non-use ($\bar{a}_1 = 0$) throughout follow-up. In both regimes of interest, we consider a hypothetical intervention to prevent censoring $\bar{c} = 0$.

(b) <u>Counterfactual outcomes of interest</u>

Our counterfactual outcomes of interest are the time to CKD under each possible longitudinal intervention to set SGLT2 use (and prevent censoring):

$$Y_{\bar{a}_1 = a, \bar{c}=0}(t), \ t = 1,..., K+1$$

Where $a \in A = \{0, 1\}$

(c) <u>Target causal parameter</u>

The primary causal estimand is the difference in the two-year, counterfactual risk of renal disease under sustained use of SGLT2 versus no use throughout follow up:

$$E[Y_{\bar{a}_1=1, \bar{c}=0}(t)] - E[Y_{\bar{a}_1=0, \bar{c}=0}(t)]$$

at time t=(K+1) corresponding to the maximum follow-up time of two years.

We are also interested in contrasting the two-year counterfactual survival curves over time. Therefore, .
we will visualize the counterfactual curves by plotting the counterfactual risk at each time point t, over time for each level of the exposure

$$E[Y_{\bar{a}_1=1, \bar{c}=0}(t)], E[Y_{\bar{a}_1=0, \bar{c}=0}(t)], \ t = 1,..., K+1$$

We will then contrast the areas under these curves by summing the causal risk differences across time:

$$\sum_{t=1}^{K+1} [E[Y_{\bar{a}_1=1, \bar{c}=0}(t)] - E[Y_{\bar{a}_1=0, \bar{c}=0}(t)]]$$

We will test this against the hypothesis that the differences are equal to zero using a variance estimate derived by the delta method.

## Observed Data

### Study population

We will collect data on a retrospective cohort of adult (age 18 years or older) patients with type II diabetes and who have initiated any second-line regimen by 2014 or later. Specifically, we will extract patient data from electronic medical records from a large integrated healthcare system in the United States, spanning across 4 sites in three different states. Patients will be excluded if they have evidence of prior CKD diagnosis. Index date for cohort entry will be defined as the date of switch to second-line regimen, among persons meeting all other inclusion criteria.

### Observed Data structure

Individual level data from this retrospective cohort can be described by a general longitudinal data structure, denoted as:

$$O = (L(0), A(0), ..., L(K), A(K), L(K+1)) \sim P_0 \in M$$

where L(0) and A(0) are baseline covariate and exposure measurements, respectively (defined explicitly below). Time is discretized into 60-day intervals, and data are collected until a maximum time of t=K+1, where K+1 is the maximum follow up time of two years or twelve 60-day intervals [i.e., K=11].

*A(t): Time-varying intervention variables, t=0,...,K*

Our intervention variables will include the time-varying exposure of use of SGLT2 inhibitors and the time-varying censoring variable:

$$A(t) = (A_1(t), C(t))$$

$A_1(t)$ takes the value of 1 when on SGLT2, and 0 when not on SGLT2. This will be operationalized as any use (from one day up to 60 days) of the corresponding drug during the interval.

$C(t)$ denotes administrative end of follow up and death. This is a time-varying variable and will take the value of 0 prior to censoring, and 1 at and after the time of censoring. Administrative censoring date will be defined as December 31, 2023. Death will also be considered a censoring event, and is, thus, also included in C(t). All-cause mortality will be measured through state death certificates.

*L₁(t): Time-varying covariates of interest, t=0,...,K+1*

Time-varying covariates will be measured at the start of each interval and defined as an indicator of presence at start of time t; each covariate takes the value of zero until evidence of the condition or medication is present in the time interval, at which point the covariate takes the value of one for the remainder of follow-up. L₁(t) variables will include diagnoses and medication covariates. These variables include: insulin use (medication NDC codes), cardiovascular disease (ICD-10 diagnosis codes, medication NDC codes; ischemic heart disease, myocardial infarction), hypertension (ICD-10 diagnosis codes, medication NDC codes).

In addition, for the baseline time point t=0, L₁(0) includes the following measured baseline covariates: age, sex, education, income, neighborhood deprivation index.

For t=1,...,K, L(t) also includes the time-varying outcome Y(t), defined below. i.e., $L(t) = (L_1(t), Y(t))$

*Y(t): Outcomes of interest (included in L(t))*

The outcome $Y$ will take the value of 0 prior to onset, and jump to 1 in the time bin immediately following that which they develop the outcome, and and remain at 1 deterministically for the remainder of follow up. We define Y(t) to be CKD diagnosis by time t, defined by <u>either</u> of the following criteria:
1) CKD diagnosis, measured through inpatient diagnosis code; or
2) glomerular filtration rate (GFR) lab results strictly less than 60 mL/min.

# Identifiability

## Sequential Randomization Assumption

$$Y_{\bar{a}=a,\bar{c}=0}(t) \perp\!\!\!\perp A(t) | \bar{L}(t), \bar{A}(t-1), t = 0, ..., K$$

Holds if for each intervention node A(t), the measured past is sufficient to block all back door paths from A(t) to Y(k) for t<k< K+1. In particular, for this application, we are particularly concerned with time invariant and time varying variables that affect the decision to modify SGLT2 and affect (or predict) CKD, as well as censoring events (death, administrative end of study).

### Positivity assumption

Need some positive probability of continuing to follow the regime of interest at each time point, given covariates:

$$P(A(t) = a(t) | \bar{A}(t-1) = \bar{a}(t-1), \bar{L}(t)) > 0, \; t = 1, ..., K \quad a.e.$$

## Statistical model and estimand

Here, our statistical model M is nonparametric (making no assumptions set of possible observed data distributions). For the final time point (*t=K+1*) and a given exposure/censoring level $\bar{a}$, our statistical estimand is the iterated conditional expectation implementation of the g-computation formula[3–5]:

$$E(Y_{\bar{a}}(K+1)) = E[E[... E[E[Y_{\bar{a}}(K+1) | \bar{L}(K), \bar{A}(K) = \bar{a}(K)]| \bar{L}(K-1), \bar{A}(K-1) = \bar{a}(K-1)] ... | L_0]]$$

Where $\bar{a}$, used for notational convenience, represents our regimes of interest and contains both treatment $\bar{a}_1$ and censoring $\bar{c}$.

Taking the difference of $E(Y_{\bar{a}}(K+1))$ with and without sustained SGLT2 use (and intervening to prevent censoring) provides our primary statistical estimand, corresponding to the two-year, counterfactual risk difference under the identifiability assumptions. Then summing the difference between the counterfactuals of interest (see section "Causal effect of interest", subsections b and c) for all time points provides our contrast of the estimated survival curves.

## Estimation

We will utilize simulations (detailed below) to select the optimal estimation method, including nuisance parameter estimation and variance estimation. To estimate our parameter of interest, we will use longitudinal TMLE[6] with Superlearner[7] as implemented via the "ltmle" R package[8].

### Simulation study implementation

Prior to analyzing the registry data and finalizing this statistical analysis plan, we conducted a simulation study to assess estimator performance, benchmark computation time, and finalize the coding of the TMLE implementation for both point estimation and inference.

We created longitudinal exposure, censoring, outcome and covariate data guided by the Observed Data description above. To generate realistic data, we sampled from the empirical data distribution of the following baseline covariates: age, sex, education, income, neighborhood deprivation index. We additionally simulated longitudinal exposure, censoring, as well as the

time-varying covariates and outcome. Time-varying covariates included insulin use, cardiovascular disease, (ischemic heart disease, myocardial infarction), and hypertension. For each time point, the past history was used to predict the current time point, consistent with our structural causal model. We utilized highly adaptive LASSO (HAL) models for the prediction.[9] For this reason, HAL models were removed from the candidate algorithm set.

We simulated 70k patients per each dataset, and created 1,000 datasets. With each simulated dataset, we used longitudinal TMLE to estimate the statistical parameter (using the longitudinal g-formula), comparing sustained SGLT2 usage to sustained non-usage over time, using 60-day time discretization and a 2-year follow-up. We simulated the truth via large-sample (n=1000000) Monte Carlo Method simulation, setting the exposure to the desired counterfactual levels of interest.

Within each estimator, we considered a broad selection of algorithms to be chosen by a discrete SuperLearner algorithm for nuisance parameter estimation: GLM, penalized regression, generalized additive models, gradient boosting models, and neural networks.

With the simulation, we compared the following specifications:

*Point estimation:*
- We evaluated the performance of the longitudinal TMLE estimator
- We evaluated nuisance parameter estimation with discrete SuperLearner with and without pre-screening for covariates
    - For algorithms without built-in covariate selection (e.g., logistic regressions and generalized additive models), we included versions adjusting for all covariates as well as versions adjusting for covariates prescreened via likelihood ratio tests to avoid overadjustment with covariates not related to the outcome.
    - By default the *ltmle* package truncates the estimated propensity scores at the bounds: [0.01,1.0]. We compared TMLEs with and without this bounding.
- We compared strategies using the negative log-likelihood loss function in SuperLearner

*Variance estimation:*
- We chose between the empirical variance of the influence curve and nonparametric bootstrap. In all cases, we will use the resulting variance estimate to create Wald-Type 95% confidence intervals.

We compared possible strategies for estimation and inference with the empirical variance across the simulations, oracle coverage (i.e. coverage obtained with the empirical variance), estimated variance, and 95% confidence coverage (i.e., coverage obtained with the estimated variance). First, we chose the strategy for point estimation that resulted in the lowest empirical variance, while maintaining nominal oracle coverage (i.e., close to 95%). Then, we selected the strategy for variance estimation that resulted in the most precise confidence intervals (i.e., smallest variance estimate), while maintaining nominal confidence interval coverage.

## Simulation study results & Pre-specified estimation

Longitudinal TMLE with SuperLearner and covariate screeners had better oracle coverage and lower empirical variance than longitudinal TMLE without covariate screening. There was no meaningful difference in the performance of the TMLEs with and without truncation of the estimated propensity scores; so we chose untruncated results. For estimating the variance of the selected TMLE (with screeners and without propensity score truncation), the empirical influence curve (IC) and nonparametric bootstrap had similar 95% confidence interval coverage; so we chose IC for computational ease.

The final estimation selection is longitudinal TMLE with a discrete SuperLearner (with candidate algorithms GLM, penalized regression, generalized additive models, gradient boosting models, and neural networks) with algorithm pre-screening and no propensity score truncation, with empirical IC for variance estimation.

## Interpretation

If identification assumptions are met, we will interpret the causal parameter both over time and as a cumulative measure (i.e. at time t=12):
- *After accounting for measured time-varying confounders, the counterfactual survival curves comparing sustained SGLT2 use vs no SGLT2 use over two years were/were not significantly different*
- *After accounting for measured time-varying confounders, two years of continuous use of SGLT2i was shown to decrease (or not) the risk of renal disease onset by X%*

## Version history

- Version 1.0: We pre-specified the simulation study to objectively select the primary analytic approach (see section, "Simulation study implementation".
- Version 1.1 (this version): We included the section, "Simulation study results & pre-specified estimation", containing results from the simulation.

# Supplementary Material B

# Example Statistical Analysis Plan (SAP) inspired by the SEARCH-Youth Study

**Laura B. Balzer**
With the SEARCH-Youth Study Team

SEARCH-Youth was a cluster (clinic) randomized trial evaluating the effect of a multicomponent intervention on HIV viral suppression among adolescents and young adults (15-24 years) with HIV in rural Uganda and Kenya (Clinicaltrials.gov: NCT03848728). The results of the trial can be found at Ruel et al.[1] with corresponding Statistical Analysis Plan (SAP) in Balzer et al.[2] In the following, **we present a simplified and hypothetical SAP**, anchored on the Causal Roadmap and inspired by the primary endpoint in the SEARCH-Youth study. We again refer the reader to [1,2] for more information about the study and the actual SAP, including additional details for the primary analysis, secondary endpoints compared between arms, implementation outcomes, descriptive statistics, and power calculations. We gratefully acknowledge the SEARCH-Youth study team.

**Roadmap Step 1: Research question**

The SEARCH-Youth trial is designed to evaluate the effect of a multi-level, combination intervention on viral suppression (HIV RNA<400 c/mL) at two-years among adolescents and young adults (15-24 years) with HIV and in care at government-sponsored health clinics in rural western Kenya and southwestern Uganda. Since the intervention involves changes in clinical practice, the unit of randomization is clinics: 14 intervention and 14 control, balanced on country. Participants randomized to intervention clinics are offered life-stage assessments, flexible care access, and rapid feedback of viral suppression status. Providers at intervention clinics have a secure WhatsApp platform for inter-provider consultation. Participants and providers at control clinics follow the standard-of-care. Our primary research question is "what is the effect of the SEARCH-Youth intervention on viral suppression (HIV RNA<400 c/mL) at two-years of follow-up?"

**Roadmap Step 2: Causal model**

At enrollment, we measured the following baseline characteristics: age, gender, education attained, marital/partnership status, alcohol use, mobility, antiretroviral therapy (ART) combination, viral suppression status, and HIV care status (recently engaged, previously engaged, or re-engaging in care). Let $W$ denote these characteristics for a given participant. Additionally, let $E^c$ denote clinic-level, baseline characteristics, such as country, the number of persons 15-24 years in HIV care, and the proportion with viral suppression among those engaged at baseline. Let $A^c$ be a clinic-level indicator of being randomized to the intervention. Throughout superscript c denotes cluster-level variables.

During endpoint data collection, we will ascertain whether the participant has outmigrated $M$, defined as moving ≥3 hours traveling distance from the study clinic. As detailed in the Study Protocol, we will attempt viral load measurement — via clinic-based ascertainment or community-based tracking — regardless of outmigration status. Let $\Delta$ be an indicator of HIV RNA (i.e., viral load) measurement at endline for a given participant. Finally, $Y$ be an indicator of viral suppression at two-years. Specifically, $Y$ is
- 1 if the participant's endline viral load is <400 c/mL
- 0 if their endline viral load is ≥400 c/mL
- otherwise missing.

Using the following hierarchical, non-parametric structural equation model,[3,4] we represent the relationships between these variables. Throughout, bold denotes matrices or vectors of the individual-level variables across participants in a given cluster.

$$E^c = f_{E^c}(U_{E^c})$$
$$\boldsymbol{W} = f_{\boldsymbol{W}}(E^c, \boldsymbol{U_W})$$
$$A^c = f_{A^c}(U_{A^c})$$
$$\boldsymbol{M} = f_{\boldsymbol{M}}(E^c, \boldsymbol{W}, A^c, \boldsymbol{U_M})$$
$$\boldsymbol{\Delta} = f_{\boldsymbol{\Delta}}(E^c, \boldsymbol{W}, A^c, \boldsymbol{U_\Delta})$$
$$\boldsymbol{Y} = f_{\boldsymbol{Y}}(E^c, \boldsymbol{W}, A^c, \boldsymbol{M}, \boldsymbol{\Delta}, \boldsymbol{U_Y})$$

By design the unmeasured factors contributing the intervention assignment $U_{A^c}$ are independent of the others. This causal model represents the data generating process for a clinic. Importantly, this causal model allows for arbitrary dependence within a cluster.

**Roadmap Step 3: Causal effect**

Let $Y_i^c(a^c)$ denote the counterfactual proportion of participants with viral suppression at two-years if possibly contrary-to-fact clinic $i$ were randomized to arm $A^c = a^c$. To increase interpretability and precision, the primary effect measure will be for the $N = 28$ clinics in the trial (i.e., the sample effect).[5,6] Specifically, we focus on the sample prevalence ratio

$$\Phi(1) \div \Phi(0) = \frac{1}{N}\sum_{i=1}^{N} Y_i^c(1) \div \frac{1}{N}\sum_{i=1}^{N} Y_i^c(0)$$

In secondary analyses, we will evaluate the sample average treatment effect (SATE): $\Phi(1) - \Phi(0)$.[5,6] In additional secondary analyses, we will include weights in these empirical means to weight individuals (instead of clinics) equally, as detailed in Benitez et al.[7]

**Roadmap Step 4: Observed data and statistical model**

The observed data for each clinic are denoted $\boldsymbol{O} = (E^c, \boldsymbol{W}, A^c, \boldsymbol{M}, \boldsymbol{\Delta}, Y)$ and are assumed to arise from a data generating process compatible with the prior causal model (Roadmap Step 2). The resulting statistical model for the set of possible observed data distributions is semi-parametric. We know that with probability 0.5 a given clinic is either intervention or control: $\mathbb{P}(A^c = 1) = 0.5$.

**Roadmap Steps 5 & 6: Identifiability, estimation, and inference**

The primary approach will be Two-Stage TMLE, which is appropriate for cluster randomized trials with missing outcomes.[8,9] In Stage-I, we first identify and estimate a cluster-level endpoint that accounts for missingness/censoring at the individual-level. Then in Stage-II, we use the endpoint estimates to evaluate the cluster-level effect with maximum precision, while accounting for limited numbers of clusters (i.e., small sample size). As detailed in Balzer et al.,[8] the Two-Stage approach lets the measurement mechanism vary by cluster and circumvents complicated estimands from the missing data equivalent of time-dependent confounding. Furthermore, the Two-Stage approach enables us to focus our Stage-II efforts on empirical efficiency maximization.

*Stage-I: Identification & Estimation*

In the primary analysis, we will include all participants and adjust for potentially differential measurement of viral suppression status.[i] To identify the cluster-level endpoint under a hypothetical intervention to ensure complete outcome measurement, we need the following missing-at-random assumption to hold within each cluster. Specifically, within values of

---

[i] Note in the actual SEARCH-Youth study, the primary endpoint considered missing endpoints to be failures (i.e., as unsuppressed); this is discussed in sensitivity analysis #2, described below.

adjustment variables, we will assume that participants with measured viral loads are representative of participants with missing viral loads *within* that cluster. Our adjustment set will consist of age, sex, baseline suppression status, and the time-varying indicator of outmigration. Therefore, our statistical estimand is

$$Y^c = \mathbb{E}[\mathbb{E}(Y|\Delta = 1, W, M)]$$

The cluster-level covariates $E^c$ and exposure $A^c$ are constant in each cluster and, thus, omitted. For the cluster-level endpoint $Y^c$ to be well-defined, we need an additional assumption on data support:

$$\mathbb{P}(\Delta = 1|W, M) > 0, \text{ a.e.}$$

In words, we need a positive probability of viral load measurement within all possible values of adjustment variables (in each cluster).

For estimation of the cluster-level endpoint $Y^c$, we will implement targeted minimum loss-based estimation (TMLE).[8,10] To reduce the risk of bias due to parametric modeling assumptions and to improve efficiency, we will use Super Learner, an ensemble machine learning method, to combine predictions from generalized linear models, generalized additive models, and the mean through 10-fold cross-validation and the non-negative least squares loss.[11] We will implement TMLE with Super Learner in each cluster separately and, thus, N=28 times to obtain cluster-level endpoints $\hat{Y}_i^c$ for each clinic $i = 1, \ldots, N$.

We will conduct two sensitivity analyses to examine the robustness of our findings:
1. Excluding participants without an endline viral load measure
2. Including all participants, but considering missing viral loads as non-suppressed

The first sensitivity analysis is sometimes called a "complete-case analysis" and relies on the missing-completely-at-random (MCAR) assumption. Under the MCAR assumption, we can estimate the clinic-level endpoint $Y^c$ with the raw proportion among those measured. The second sensitivity analysis considers a composite outcome and is not subject to missingness. Therefore, we can again estimate the clinic-level endpoint $Y^c$ with the proportion with viral suppression (among all participants).

*Stage-II: Identification and Estimation*

Given the estimated cluster-level endpoints $\hat{Y}_i^c$ for $i = 1, \ldots, N$ from Stage-I, we will then evaluate the intervention effect with a cluster-level TMLE. At the cluster-level, there is no confounding or missingness. Therefore, we can easily identify and estimate the causal effect of interest (Roadmap Step 3) by contrasting the average of the cluster-level endpoints between arms. This unadjusted prevalence ratio will be used in a sensitivity analysis.

In the primary analysis, we will adjust for baseline covariates to improve efficiency, as repeatedly demonstrated.[12–20] In other words, our statistical estimand for Stage-II is given by

$$\Psi(1) \div \Psi(0) = \frac{1}{N}\sum_{i=1}^{N} \mathbb{E}(Y^c|A^c = 1, E^c{}_i, W^c{}_i) \div \frac{1}{N}\sum_{i=1}^{N} \mathbb{E}(Y^c|A^c = 0, E^c{}_i, W^c{}_i)$$

where $W^c{}_i$ denote aggregates of individual-level baseline covariates. We will use TMLE with Adaptive Pre-specification (APS) for estimation and inference. APS is a formal procedure to data-adaptively choose from a pre-specified set the adjustment variables that maximize

empirical efficiency, while protecting Type-I error control.[7–9,21,22] Our candidate adjustment variables will be the clinic-specific number of adults 15-24 years in HIV care at baseline, the clinic-specific proportion with viral suppression among those engaged at baseline, or nothing (unadjusted). Our candidate estimation algorithms for the outcome regression $\mathbb{E}(Y^c|A^c, E^c, W^c)$ will be limited to working GLMs[16] adjusting for at most of covariate. Our candidate estimation algorithms for the known propensity score $\mathbb{P}(A^c = 1|E^c, W^c)$ will also be limited to working GLMs adjusting for at most of covariate. Using leave-one-out cross-validation, we will then select the combination of estimators and, thus, the TMLE that minimizes the cross-validated variance estimate.

*Statistical Inference*

Under the conditions detailed in Balzer et al.,[8] Two-Stage TMLE will be an asymptotically linear estimator, meaning that it will be normally distributed in the large data limit. Informally, these conditions require that the cluster-level endpoints are estimated well in Stage-I. For variance estimation, we use the sample variance of the estimated, cluster-level, influence curve. As a finite sample approximation to the standard normal distribution, we will use the Student's t-distribution with 28-2=26 degrees of freedom.[12]

Using a two-sided test at the 5% significance level,[ii] we will formally test the *null hypothesis* that the SEARCH-Youth intervention did not impact viral suppression, as compared to the standard-of-care. We will report point estimates and 95% confidence intervals for the arm-specific endpoints as well as the relative and absolute effects.

## Roadmap Step 7: Interpretation

We will interpret the effect analyses in light of our missing data assumptions and the randomization of the intervention. In the actual SEARCH-Youth Study, Ruel et al. concluded, "The mean proportion of participants with virological suppression at 2 years was 88% (95% CI 85–92) for participants in intervention clinics and 80% (77–84) for participants in control clinics, equivalent to a 10% beneficial effect of the intervention (risk ratio [RR] 1.10, 95% CI 1.03–1.16; p=0.0019)."[1]

---

[ii] Note in the actual SEARCH-Youth study, the pre-specified hypothesis test was one-sided, corresponding to the null hypothesis of no improvement in viral suppression due to the intervention.

## How simulations informed this SAP

Prior to unblinding the data, we pre-specified and conducted the following treatment-blind simulations to select the primary analytic approach that strictly controlled Type-I error. As a secondary criterion, we evaluated 95% confidence interval coverage for the cluster-level endpoints $Y^c$.

*Treatment-blind simulations with dependent and missing outcomes*

Recall the individual-level outcome $Y$ is subject to missingness but the true measurement mechanism is unknown. To evaluate estimator performance with simulations, we re-defined the outcome $Y$ to be 1 if the participant's endline viral load was <400 c/mL and 0 otherwise (including endline viral load ≥400 c/mL or missing). Then we generated 1000 plasmode datasets as follows. We preserved the real covariate and (redefined) outcome data, but randomly permuted the treatment indicator $A^c$ across clusters. Then an independent statistician simulated individual-level, measurement indicators $\Delta$ as a function of the baseline cluster-level and individual-level covariates, the permuted exposure, and time-varying covariates. The primary statistician (LBB) was blinded to the functional form of the measurement mechanism. In the simulation, the (redefined) outcomes were set to missing if the simulated measurement indicator $\Delta$ was 0. Using these simulated datasets, we compared various implementations of Two-Stage TMLE, as follows.

<u>First, we focused on Stage-I estimation</u> of the cluster-level endpoint $Y^c = \mathbb{E}[\mathbb{E}(Y|\Delta = 1, W, M)]$ and compared
- The empirical mean (i.e., the unadjusted proportion with the outcome among those measured)
- TMLE with main terms regression for nuisance parameter estimation
- TMLE with Super Learner (combining predictions from generalized linear models, generalized additive models, and the mean with 10-fold cross-validation and non-negative least squares loss) for nuisance parameter estimation

For all, we fixed our approaches to Stage-2 estimation to the unadjusted effect estimator and for variance estimation to the standard, cluster-level influence curve (IC). This allowed us to evaluate the potential for alternative Stage-I approaches (both identifiability assumptions and estimators) to reduce bias due to differential outcome measurement.

To evaluate these approaches, we calculated the proportion of simulated datasets where the 95% confidence interval excluded the null (equivalent to the Type I error rate). As expected, only flexible adjustment for missing outcomes using TMLE with Super Learner in Stage-I preserved the Type-I error rate at $\leq 5\%$ (Table 1). Additionally, in these simulations the true values of the cluster-level endpoints $Y^c$ in the absence of missingness were known; therefore, we also verified this approach resulted in nominal-conservative 95% confidence interval coverage for the cluster-level endpoints.

<u>Second, we focused on Stage-II estimation</u> of the intervention effect $\Psi(1) \div \Psi(0)$ and compared

- TMLE with "limited APS" where candidate estimators were limited to working GLMs adjusting for at most a single covariate[21]
- TMLE with "expanded APS" where candidates including worked GLMs adjusting for a single covariate, working GLMs adjusting for multiple covariates, penalized regression, and the unadjusted estimator[22]
- Standard versus cross-validated estimates of the cluster-level IC for variance estimation[21]

For all, we fixed our approach to Stage-I estimation of the cluster-level endpoints as TMLE with Super Learner. This allowed us to evaluate the potential efficiency gains through data-adaptive adjustment for baseline covariates in Stage-II, after appropriately addressing missing data in Stage-I. As shown in Table 2, the limited APS implementation in Stage-II and the standard IC approach to variance estimation resulted in the best Type-I error control.

Thus, we settled on the following implementation of Two-Stage TMLE as the primary analytic approach: Stage-I estimation using TMLE with Super Learner, Stage-II estimation with TMLE using limited APS, and variance estimation with the standard IC approach. Two-Stage TMLE with the unadjusted effect estimator in Stage-II was a pre-specified secondary analysis.

| Stage-I estimation | Stage-II estimation | Variance estimation | Type-I error rate for $\Psi(1) \div \Psi(0)$ | 95% CI coverage for $Y^c$ |
|---|---|---|---|---|
| Simple mean | Unadjusted | Standard IC | 11% | 80% |
| TMLE with main terms regression | Unadjusted | Standard IC | 8% | 83% |
| TMLE with Super Learner | Unadjusted | Standard IC | 3% | 96% |

*Table 1:* Hypothetical simulation results when comparing Stage-I estimators of the cluster-level endpoint $Y^c$. All approaches used the unadjusted effect estimator in Stage-II and the standard influence curve (IC)-based inference.

| Stage-I estimation | Stage-II estimation | Variance estimation | Type-I error rate for $\Psi(1) \div \Psi(0)$ |
|---|---|---|---|
| TMLE with Super Learner | Limited APS | Standard IC | 5% |
| TMLE with Super Learner | Limited APS | Cross-validated IC | 2% |
| TMLE with Super Learner | Expanded APS | Standard IC | 10% |
| TMLE with Super Learner | Expanded APS | Cross-validated IC | 7% |

*Table 2:* Hypothetical simulation results when comparing Stage-II estimators of the intervention effect $\Psi(1) \div \Psi(0)$ and approaches to variance estimation. All approaches used TMLE with Super Learner in Stage-I.

## History of changes

- Version 1.0: We pre-specified the simulation study to objectively select the primary analytic approach: data generating processes, candidate estimators, and estimation metrics.
- Version 2.0 (this version): We included the results of the simulation study and pre-specified the primary and secondary/sensitivity analyses.

## Acknowledgements



%TC:endignore
\end{document}